\documentclass[12pt,a4paper]{article}
\usepackage{epsfig,graphicx}
\newcommand{\be}{\begin{equation}}
\newcommand{\ee}{\end{equation}}
\newcommand{\bea}{\begin{eqnarray}}
\newcommand{\eea}{\end{eqnarray}}
\newcommand{\beas}{\begin{eqnarray*}}
\newcommand{\eeas}{\end{eqnarray*}}
\newcommand{\ba}{\begin{array}}
\newcommand{\ea}{\end{array}}
\newcommand{\nn}{\nonumber}

\newcommand{\bt}{\begin{table}}

\newcommand{\vsi}{\varsigma}

\newcommand{\de}{\delta}

\newcommand{\ka}{\kappa}
\newcommand{\la}{\lambda}
\newcommand{\La}{\Lambda}
\newcommand{\na}{\nabla}

\newcommand{\si}{\sigma}

\begin{document}
\title{\bf\noindent
Scalar graviton  as dark matter}
\author{Yu.~F.~Pirogov
\\
\small{\em Theory Division, Institute for High Energy Physics,  Protvino, 
Moscow Region, Russia }
}
\date{}
\maketitle

\begin{abstract}
\noindent
In the report, the theory of unimodular bimode gravity
built on principles of unimodular gauge invariance/relativity and general
covariance is  exposed.   Besides  the massless tensor graviton of
General Relativity, the theory includes  an (almost) massless scalar graviton 
treated  as the gravitational dark matter.  A spherically
symmetric vacuum solution,  describing  the coherent scalar-graviton   field
for  the soft-core dark halos with   the asymptotically flat
rotation curves, is demonstrated.
\end{abstract}

\section*{Introduction}
The report presents  an extension to General Relativity (GR) based  on  the
unimodular relativity and general covariance~\cite{Pir0}--\cite{Pir1}, 
the so-called Unimodular Bimode Gravity (UBG).\footnote{For  a self-contained  
exposition, see~\cite{Pir1}.}
The basic principles of the theory  and the simplified Lagrangian model  are
indicated. The dark  halos  built of  a  new 
particle -- the (nearly) massless scalar graviton  -- are shown
to naturally emulate  the galaxy dark matter (DM) halos.

\section{Unimodular relativity}

\paragraph{Gauge invariance/relativity}

To be  consistent as  the  effective field theory, in particular, to allow for
the  quantum corrections,  a gravity theory, describing the spin-two field,  
should be built on a gauge principle.
The essence of the  theory is determined  by  its  gauge properties 
under  the diffeomorphism transformations,  and the group of  the gauge 
invariance/relativity.  More particularly, the diffeomorphisms are given by the
coordinate  transformations:   $x^\mu\to x'^\mu =  x'^\mu(x)$ followed by the
field substitutions  $\varphi (x)\to \varphi'(x')$. 
At that, the local properties if the theory are determined by the infinitesimal
diffeomorphims: $x^\mu\to x'^\mu= x^\mu    + \de_\xi x^\mu$,   with $ \de_\xi
x^\mu\equiv -\xi^\mu$ being a vector field.  These properties are
expressed through the  so-called Lie derivative  given by the net infinitesimal 
variation of a field exclusively due to its tensor properties: $\delta_\xi 
\varphi(x) = (\varphi'(x')-\varphi(x'))_{x'\to x}$. 
In particular,  one has for metric
\bea
\delta_\xi g_{\mu\nu}&=& g_{\mu\la} \partial_\nu
\xi^\la+  g_{\nu\la}  \partial^\la\xi_\mu+ g_{\mu\nu}  
\partial_\la\xi^\la    \nn\\
&=& \na_\mu \xi_\nu+ \na_\nu \xi_\mu,
\eea
where $\xi_\mu= g_{\mu\nu}\xi^\nu$ and $\na_\mu$ is a covariant derivative.
For a scalar field one has 
$
\de_\xi \phi= \xi^\mu \partial_\mu \phi.
$
The general gauge  invariance/relativity corresponds to 
the group of the general diffeomorphisms (GDiff):
\be
\mbox{\rm GDiff}: \   \xi^\mu \ \mbox{\rm unrestricted}.
\ee
By means of GDiff, one  can eliminate in metric all the
excessive degrees  of freedom, but for two corresponding to  the
transverse-tensor graviton. As a result, 
this ensures the masslessness of the graviton, $m_g=0$. This  takes place in
GR, as well as in its generally invariant extensions.

Of particular interest is  the field  $\sqrt{-g}$, $g\equiv\mbox{\rm
det}(g_{\mu\nu}) $,  which may, under certain  conditions,  present an
additional,
scalar degree of freedom  in metric. The respective  Lie derivative is
\be
 \delta_\xi\sqrt{-g}   
=\frac{1}{2}\sqrt{-g}g^{\mu\nu}\de_\xi g_{\mu\nu} 
= \partial_\mu( \sqrt{-g} \xi^\mu ),
\ee
or otherwise
\be
 \delta_\xi \ln \sqrt{-g} =  \xi^\mu \partial_\mu \ln  \sqrt{-g}
+\partial_\mu \xi^\mu.
\ee
Define  the group of the  transverse diffeomorphisms (TDiff) as
\be
\mbox{\rm TDiff}: \   \partial_\mu \xi^\mu=0.
\ee
It proves,  that TDiff is necessary and sufficient for $m_g=0$~\cite{Ng}, with
GDiff of  GR being, in fact, overabundant for this purpose. Moreover,  it
is seen   that under TDiff the quantity 
$\ln \sqrt{-g}$ behaves like a scalar field and may thus 
be used   as an extra degree of freedom in metric, not spoiling the
masslessness of the (tensor) graviton. 
The metric theory of gravity  built on the transverse gauge
invariance/relativity  is the so-called Transverse Gravity
(TG) (see, e.g.,~\cite{Alvarez}). However, at face value, such a theory is not
generally covariant.
To remedy this  introduce a nondynamical scalar
density $ \bar\mu$ (the  so-called modulus) transforming as $\sqrt{-g}$, so that
$\vsi\equiv \ln
(\sqrt{-g}/\bar\mu)$ behaves like a scalar field under GDiff. 
Define  then  the group of the unimodular diffeomorphisms (UDiff) as preserving
the modulus:
\be
 \mbox{\rm UDiff}: \  \de_\xi \bar\mu  = \partial_\mu(\bar\mu \xi^\mu)=0, 
\ee
The theory of gravity built on the  unimodular
invariance/relativity  is   Unimodular Gravity (UG). 
Being generally
covariant, UG is equivalent  to the non generally covariant TG in coordinates
where $\bar \mu=1$ (if any).  
But now due to the general covariance,  
one can use GDiff to study the theory in the arbitrary observer's coordinates.

\paragraph{Physical/helicity  content}

Indicate   the physical/helicity   content of UG. 
Consider the weak-gravity-field
approximation:
$g_{\mu\nu}=\eta_{\mu\nu}+ h_{\mu\nu}$, $\sqrt{-g } =1+h/2$, with
$h=\eta^{\mu\nu}h_{\mu\nu}=h_{00}-h_{ll}$. 
Under GDiff, one has
\bea
\de_\xi h&=&\xi^\mu\partial_\mu h+2 \partial_\mu \xi^\mu,\nn\\
\de_\xi \bar h&=&\xi^\mu\partial_\mu\bar  h+2 \partial_\mu \xi^\mu,
\eea
where  $\bar h/2\equiv\ln\bar\mu$ (not necessarily small).
This means that $h$ and $\bar h$ separately  can be eliminated  due to 
GDiff.  Nevertheless, $\vsi=(h-\bar h)/2$ transforming homogeneously
as a scalar field  can  not be removed  and may thus 
serve as an independent
degree of freedom. 
With the rest of components in $h_{\mu\nu}$ transforming under GDiff 
exactly as in GR,  the  physical/helicity  content of UG thus includes:\\
--  transverse deformation (4-volume preserving)/tensor
graviton/{\em graviton} ($\la_g=\pm 2$);\\ 
-- compression (form preserving) mode/scalar graviton/{\em systolon}
($\la_s =0$).\\
The bimode gravity  mediated  by the   systolon and graviton   may naturally be
referred to as  the  {\em systo-gravity}.

\section{General covariance}

\paragraph{Unimodular Gravity} 

To be precise, under the Unimodular Gravity (UG) we  understand   collectively
any generally covariant   and unimodular invariant   metric theory  of gravity
containing a nondynamical scalar density (modulus) $\bar\mu$.
For general covariance, $\bar\mu$ may enter only through  $\bar\mu/\sqrt{-g}$.
The UG action in the vacuum looks generically like:
\be
S=   \int{ \cal L}\, d^4 x   =  \int  L   (g_{\mu\nu},\bar\mu /\sqrt{-g} ) 
\sqrt{-g} \,d^4 x,
\ee
with $ L$ being the scalar  Lagrangian.
The  UG  field equations (FEs) in vacuum are obtained by extremizing $S$ under 
$\de \bar\mu =0$, with the result
\be
{\cal G}^{\mu\nu}\equiv \delta {\cal L}/ \de  g_{\mu\nu}=0, 
\ee
and $\bar\mu$ appearing as an external  functional parameter.
Ultimately, this signifies  the openness of the system of fields considered and
may, conceivably, serve as an off-spring of a ``new physics''.
Some particular cases of UG are as follows.

\paragraph{General Relativity}   

The marginal case of with the absence of 
$\bar\mu$ corresponds to GR:
\bea
{\cal L}&=&-\ka_g^2\Big( \frac{1}{2} R -\La\Big) \sqrt{-g},\nn\\
 &&R^{\mu\nu}-\frac{1}{2}R g^{\mu\nu }  +\La g^{\mu\nu }=0,
\eea
where $\ka_g$ is the Planck mass and $\La$ is a cosmological constant. 
The  GR vacuum corresponds thus  to $ R= 4\La$.

\paragraph{Unimodular  Relativity} 

The restriction $\sqrt{-g}/\bar\mu=1$
corresponds to  the
so-called Unimodular Relativity (UR)~\cite{Anderson}:
\bea
{\cal L}&=  &-\frac{\ka_g^2}{2}(R\sqrt{-g})|_{\sqrt{-g}=\bar\mu} ,\nn\\
&& R^{\mu\nu}-\frac{1}{4} Rg^{\mu\nu}=0  \  ,
\eea
so that  $R$ remains undetermined by  FEs.  The Bianchi identity  gives
additionally $ \partial_\mu R=0$. Henceforth,  the UR vacuum corresponds to 
$R=\La_0$, 
with $\La_0$ being an  integration constant, reproducing effectively the 
Lagrangian $\La$-term.  Thus UR is classically equivalent to GR with a
cosmological constant, both theories describing only massles tensor graviton.

\paragraph{Unimodular  Bimode Gravity} 

The radically new gravity theory appears when $\bar\mu/\sqrt{-g}$  enters
 through the additional kinetic term. This case corresponds to Unimodular 
Bimode Gravity (UBG/systo-gravity)~\cite{Pir1}:
\bea
{\cal L}&=& \Big(-\frac{\ka_g^2}{2} R + \frac{\ka_s^2}{2}
g^{\mu\nu}\partial_\mu
\vsi \partial_\nu \vsi\Big) \sqrt{-g},\nn\\
\vsi&\equiv  &\ln \sqrt{-g}/\bar\mu.  
\eea
In addition to the massless transverse-tensor graviton, UBG/systo-gravity
describes
the (nearly) massless scalar graviton/systolon $\vsi$, characterized by  the
appropriate mass scale $\ka_s<\ka_g$.
The UBG/systo-gravity Lagrangian is unique because:\\ 
-- the  higher derivative terms with 
$\vsi$ are suppressed by  the powers of $1/\ka_s$;\\ 
-- the  derivativeless terms with~$\vsi$ are forbidden  
under the global symmetry $ \vsi \to \vsi+\vsi_0$.\footnote{For 
potential $V_s(\vsi)$ violating the global  symmetry and producing, in
particular, the  mass for $\vsi$,  see~\cite{Pir2, Pir1}.}\\
The FEs for UBG/systo-gravity are
\be
R^{\mu\nu}-\frac{1}{2}g^{\mu\nu}  
-\ka_g^{-2}T_s^{\mu\nu}=0, 
\ee
where $T_{s\mu\nu}$ is the effective  systolon energy-momentum tensor: 
\be
T_{s\mu\nu}=\Big( \partial_\mu \si\partial_\nu \si   -   \frac{1}{2}
\partial \si \cdot \partial \si g_{\mu\nu}\Big)    + \ka_s \na\cdot
\na \si g_{\mu\nu} ,
\ee
where $ \si\equiv \ka_s \vsi$,
The  non-harmonic  term  $\ka_s\na\cdot \na \si$ is specific to
UBG/systo-gravity and may serve  as  a signature of the latter, being absent in 
other gravity  theories  such~as, e.g.:\\ 
-- GR with a free massless scalar field;\\
--  the free Brans-Dicke  theory in Einstein frame.\\
In essence, such a  term,   ensuring  a  peculiar  dark halo solution, is  in
charge for treating the systolon as a DM particle.

\section{Dark halos}

\paragraph{Harmonic scalar field} 

 In the case   $\na\cdot\na\si =0$, 
the UBG/systo-gravity FEs  are  similar to those of GR
with a free massless scalar field. Having singularity at the origin,  their
solutions  describe the so-called dark holes/{\em fractures} consisting of the
scalar field $\si$ and the matter at the origin.
The particular cases of the dark holes/fractures  are:\\
--  the black holes: no scalar field;\\
-- the vacuum holes: no matter.\footnote{For  more detail on
the dark holes/fractures, in particular,  on the difference between their
physical interpretation in UBG/systo-gravity and GR with a scalar field,
see~\cite{Pir1', Pir1}.} 

\paragraph{Non-harmonic scalar  field} 

In the case $\na\cdot\na\si\neq 0$,
there appears a principally new spherically symmetric regular solution to
the vacuum FEs. It looks
approximately like (Fig.~1):
\be
{\vsi}_h= 
\cases{
\tau^2 -\frac{ 3}{10} \tau^4+ {\cal O}(\tau^6), \   \tau\leq 1, \cr
\ln 3\tau^2, \hspace{13ex}   \tau\gg 1, }
\ee
where $ \tau=r/R_0$ is the scaled radial distance, with $R_0$ being an
integration
constant. This solution describes the dark halos, with $R_0$ being their
characteristic distance scales.
\begin{figure}
\begin{center}
 \resizebox{0.5\textwidth}{!}{
\includegraphics{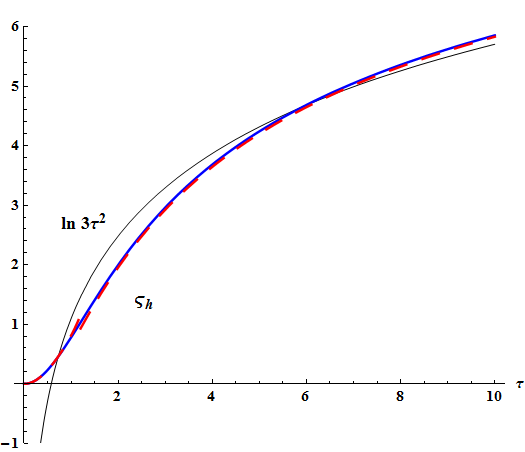}}
\end{center}
\caption{Dark halo solution $\vsi_h(\tau)$.
}
\end{figure}
The energy density profile of a  dark halo is as follows (Fig.~2):
\be
\rho_{h}/  \rho_{0} =
\cases{
 1-\tau^2+\frac{4}{5} \tau^4 
+{\cal O}(\tau^6), \hspace{1ex}   \tau\leq 1, \cr
1/(3\tau^2)  +{\cal O}(1/\tau^{5/2}) ,    \ \ \   \tau\gg 1,}
\ee
with the central energy density $ \rho_{0}= 3 \ka_s^2/R_{0}^2$.
This profile lies somewhere in between the two commonly accepted Monte Carlo
cases:\\
-- the pseudo-isothermal sphere: $\rho_{ref}/\rho_0= 1/(1+\tau^2)$ (dotted
line);\\
-- the isothermal sphere: $\bar \rho/\rho_0=1/(3\tau^2)$  (thin line).\\
In reality, there should also be included  corrections to the halo 
core due to:\\
--  luminous matter\  $\rho_{lm}$;\\
--  particle DM  $\rho_{dm}$.\\
The halo rotation curve   profile is as follows (Fig.~3):
\be
v_h^2/\upsilon_\infty^2 =\cases{
\tau^2 -\frac{3}{5}\tau^4 
+ {\cal O}(\tau^6), \   \tau \leq 1,\cr
1+{\cal O}(1/\sqrt{\tau}), \hspace{4ex} \   \tau \gg 1,}
\ee
with  $\upsilon_\infty$ being  the asymptotic rotation velocity: 
$
\upsilon_\infty= \ka_s/ \sqrt{2} \ka_g.
$
In order  for the dark halos to have baring   to  galaxies,   $\upsilon_\infty$
should be about $10^{-3}$. This approximately 
fixes   the mass scale for the scalar gravity relative to the tensor one.

\begin{figure}
\begin{center}
\resizebox{0.5\textwidth}{!}{%
\includegraphics{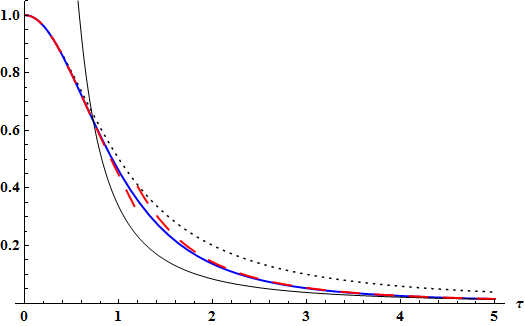}}
\end{center}
\caption{
\label{fig:6a}  
Normalized energy density profile $\rho_h(\tau)/\rho_{0}$.
}
\end{figure}

\begin{figure}   
\begin{center}
\resizebox{0.5\textwidth}{!}{%
\includegraphics{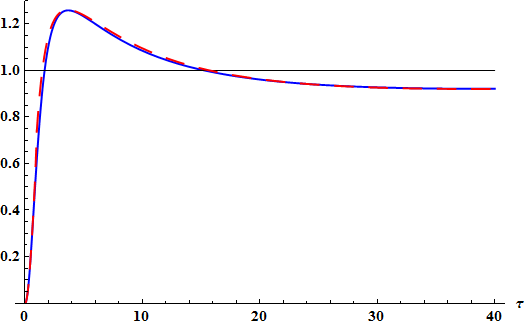}}
\end{center}
\caption{
\label{fig:8}    
Normalized rotation curve profile $v_h(\tau)/\upsilon_\infty$.
}
\end{figure}

Altogether,  one may say about  the coherent dark  halos that:\\
-- in  the vacuum,  they  are 
not only possible  but are, in a sense, ``predictable'', being  thus 
a signature of UBG/systo-gravity (similar to BHs being a signature of GR and
its extensions);\\
-- in the presence of matter,  they are likely to reflect only the universal 
long-range  tail,  with the specific short-range core influenced    by
matter (including, possibly, the particle DM  besides the luminous one);\\
-- the two-component   DM (the coherent
systolon  field    plus  the continuous medium) may  eventually present  the
long-waited solution to the DM puzzle.\footnote{For
more detail on the dark
halos, see~\cite{Pir2}--\cite{Pir1}. Besides, see~\cite{Pir1} for  the
interpolating  dark hole-halo solutions describing the so-called dark
{\em lacunas}, which may model the galaxies in toto (but for the distributed
matter).}

\section*{Conclusion} 

We conclude as follows:\\
--  UBG/systo-gravity  is the  theoretically  viable extension to GR,  safely 
preserving its solid features, as well as predicting a lot of new phenomena
beyond GR;\\
--  the emerging new particle -- the (almost) massless scalar
graviton/systolon -- reveals the properties imminent to DM;\\
-- further studying  the theoretical and observational aspects of the theory, 
especially due to inclusion of  matter, is quite urgent.


\begin{thebibliography}{*} 

\bibitem{Pir0} 
Yu.\ F.\ Pirogov, Phys.\ Atom.\ Nucl.\  {\bf 69},  1338  (2006);
arXiv:gr-qc/0505031.

\bibitem{Pir0'} 
Yu.\ F.\ Pirogov,  arXiv:gr-qc/0609103.

\bibitem{Pir1'}
Yu.\ F.\ Pirogov,  Phys.\ Atom.\ Nucl.\  {\bf 73}, 134  (2010);
arXiv:0903.2018[gr-qc].

\bibitem{Pir2}
Yu.\ F.\ Pirogov, Mod.\ Phys.\ Lett.\ A {\bf 24}, 3239  (2009); 
arXiv:0909.3311 [gr-qc].

\bibitem{Pir3}
Yu.\ F.\ Pirogov and I.~Yu.~Polev, arXiv:1010.3431 [gr-qc].

\bibitem{Pir1}
Yu.\ F.\ Pirogov, Eur.\ Phys.\ J.\ C {\bf 72}   (2012) 2017; 
arXiv:1111.1437 [gr-qc].

\bibitem{Ng}
J.~J.~van der Bij, H.\ van Dam and Y.\ J.\ Ng, Physica A {\bf 116}, 307 (1982).

\bibitem{Alvarez}
E.\   Alvarez {\em et.\ al.}, Nucl.\ Phys.\ B {\bf 756}, 148  (2006); 
arXiv:hep-th/0606019.

\bibitem{Anderson}
J.\ L.\ Anderson and D.\ R.\ Finkelstein,  Am.\  J.\ Phys.\ {\bf 39}, 901
(1971).

\end{thebibliography}
\end{document}